\documentclass[prb,showpacs,superscriptaddress,twocolumn]{revtex4-1}
\usepackage{amssymb,amsmath,color,graphicx,psfrag}

\begin{document}

\newcommand{\kv}[0]{\mathbf{k}}
\newcommand{\Rv}[0]{\mathbf{R}}
\newcommand{\rv}[0]{\mathbf{r}}
\newcommand{\K}[0]{\mathbf{K}}
\newcommand{\Kp}[0]{\mathbf{K'}}
\newcommand{\dkv}[0]{\delta\kv}
\newcommand{\dkx}[0]{\delta k_{x}}
\newcommand{\dky}[0]{\delta k_{y}}
\newcommand{\dk}[0]{\delta k}
\newcommand{\cv}[0]{\mathbf{c}}
\newcommand{\qv}[0]{\mathbf{q}}
\newcommand{\pnt}[1]{\psfrag{#1}{\tiny{$#1$}}}

\newcommand{\jav}[1]{#1}

\title{Optimal protocols for finite-duration quantum quenches in the Luttinger model}
\author{\'Ad\'am B\'acsi}
\email{bacsi.adam@sze.hu}
\affiliation{MTA-BME Lend\"ulet Topology and Correlation Research Group,
Budapest University of Technology and Economics, 1521 Budapest, Hungary}
\affiliation{Department of Mathematics and Computational Sciences, Sz\'echenyi Istv\'an University, Gy\H or, Hungary}
\author{Masudul Haque}
\affiliation{Department of Theoretical Physics, Maynooth University, Co. Kildare, Ireland}
\affiliation{Max Planck Institute for the Physics of Complex Systems, N\"{o}thnitzer Str. 38, 01187 Dresden}
\author{Bal\'azs D\'ora}
\affiliation{MTA-BME Lend\"ulet Topology and Correlation Research Group,
Budapest University of Technology and Economics, 1521 Budapest, Hungary}
\affiliation{Department of Theoretical Physics, Budapest University of Technology and Economics, Budapest, Hungary}
\date{\today}

\begin{abstract}

Reaching a target quantum state from an initial state within a finite temporal window is a
challenging problem due to non-adiabaticity.  We study the optimal protocol for swithcing on
interactions to reach the ground state of a weakly interacting Luttinger liquid within a finite time
$\tau$, starting from the non-interacting ground state.  The protocol is optimized by minimizing the
excess energy at the end of the quench, or by maximizing the overlap with the interacting ground
state.  We find that the optimal protocol is symmetric with respect to $\tau/2$, and can be
expressed as a functional of the occupation numbers of the bosonic modes in the final state.  For
short quench durations, the optimal protocol exhibits fast oscillation and excites high energy
modes.  In the limit of large $\tau$, minimizing energy requires a smooth protocol while maximizing
overlap requires a linear quench protocol.  
In this limit, the minimal energy and maximal overlap are both universal functions of the system size and the duration of the protocol.

\end{abstract}

\maketitle

\section{Introduction}

Progress in quantum technologies relies on our ability to manipulate quantum states, in particular
interacting many-component quantum states.  A key challenge is to engineer the transfer of a quantum
system from one ground state to another, without excitations, in finite time.  Such a transfer is
guaranteed by the adiabatic theorem if the duration of parameter change is allowed to be infinite.
When this is performed in finite time, this is often referred to as a `shortcut to adiabaticity'.
Such techniques are an obvious route to improving the viability of quantum annealing and adiabatic
quantum computing algorithms, \cite{Nishimori_QuantumAnnealing_PRE1998,Aeppli_Science1999,
  Farhi_adiabaticQC_2000} for which unwanted excitations are of serious concern.

The problem of optimizing a finite-duration quantum quench has been addressed in the context of a
variety of quantum systems, including
trapped particles or trapped Bose-Einstein condensates,
\cite{DelCampo_Muga_PRL2010,GueryOdelin_Muga_PRA2011, DelCampo_PRA2011,DelCampo_PRL2013}
trapped interacting fermionic gases, \cite{Rigol_adiabatic_PRE2017, delCampo_unitaryFermi_PRA2018,
  Nielsen_trappedfermions_PRA2018}
Luttinger liquids, \cite{rahmani_prl2011} 
Majorana qubits, \cite{Rahmani_vonOppen_Refael_Majorana_PRB2015, rahmani_Majorana_PRB2017}
the Lipkin-Meshkov-Glick model, \cite{Caneva_Calarco_Fazio_PRA2011, Takahashi_PRE2013, DeChiara_Fazio_LMG_PRL2015}
and spin systems. \cite{Berry_JPA2009, Caneva_Calarco_Fazio_optimal_PRL2009, DelCampo_Zurek_PRL2012, Takahashi_PRE2013, Joynt_Vavilov_PRA2018}
Optimal protocols have been studied in quantum quenches through a quantum critical point
\cite{polkovnikov2008, Caneva_Calarco_Fazio_PRA2011, DelCampo_Zurek_PRL2012,
  DelCampo_Sengupta_EPJ2015} and from a quantum critical point to the gapless phase of the Luttinger
liquid. \cite{rahmani_prl2011}

In this work, we consider the optimization of finite-duration ramps in a Luttinger liquid.
Luttinger liquids appear as effective low-energy descriptions of gapless phases in various
one-dimensional (1D) interacting systems. \cite{book_giamarchi, book_nersesyan,
  CazalillaGiamarchiRigol_RMP2011, Schoenhammer_LLreview_JPCM2012} For example, for fermions in 1D,
Landau's Fermi liquid description breaks down for any finite interaction --- the low-energy physics
is described by bosonic collective modes with linear dispersion and is characterized by anomalous
non-integer power-law dependences of correlation functions.  The Luttinger model similarly arises as
the low-energy description of spin chains or that of interacting 1D bosons. \cite{book_giamarchi,
  book_nersesyan, CazalillaGiamarchiRigol_RMP2011} In addition to its rich history in equilibrium
condensed matter physics, in the past dozen years the Luttinger model has also been used as a model
system for non-equilibrium phenomena.  Non-equilibrium studies using the Luttinger model include
investigations of
instantaneous quantum quenches,
\cite{Cazalilla_LLquench_PRL2006,Uhrig_LLquench_PRA2009, Iucci_Cazalilla_LLquench_PRA2009,
  BarmettlerGritsevDemlerAltman_XXZ_NJP2010, MitraGiamarchi_PRL2011, Mitra_quench_PRL2012,
  KarraschSchurichtMeden_LLquench_PRL2012, Schuricht_Meden_LLquench_NJP2012, Meden_LLquench_PRB2013,
  Mitra_prethermalized_PRB2013, BagretsMirlin_LLquench_PRB2013, Becca_Parola_1Dquench_EPJB2013,
  Meden_LLquench_PRL2014, SchiroMitra_LL_OrthogonalityCatastrophe_PRL2014,
  Cazalilla_Chung_LLquench_review_2016, Sassetti_LLquench_Scipost2018}
transport due to inhomogeneous initial conditions, \cite{Gutman_Gefen_Mirlin_PRB2010,
  LancasterMitra_PRE2010, Perfetto_Stefanucci_LLtransport_PRL2010,
  Protopopov_Gutman_Schmitteckert_Mirlin_PRB2013, SchiroMitra_transport_PRB2015,
  Lebowitz_LL_inhomquench_PRB2017, Dubail_Stephan_Calabrese_inhomogeneousLL_SciPost2017}
and, most relevantly to the present work,
finite-duration (finite-rate) quenches.  \cite{balazs_crossover, Dziarmaga_LL_PRB2011,
  Perfetto_Stefanucci_LL_EPL2011, Pollmann_Haque_Dora_LL_XXZ_PRB2013, AmitDutta_LLramp_PRB2014,
  BernierKollath_LLramp_PRL2014, Schuricht_LLramp_PRB2016, Sassetti_LLramp_PRB2016}

In the present paper, we consider quenches having a certain duration $\tau$, governed by a quench shape function $Q(t)$
such that $Q(0)=0$ and $Q(t>\tau)=1$.  The system starts at $t=0$ in the ground state of the initial
non-interacting Hamiltonian.  To proceed analytically, we assume a weak final interaction, which
allows for a perturbative, analytical treatment of the ensuing Bogoliubov equations.  In general,
for finite $\tau$ the final state after the quench differs from the ground state of the final
Hamiltonian.  The deviation can be quantified either by the excess energy of the final state
relative to the target ground state, or by the overlap between the final state and the target ground
state, i.e., the vacuum-to-vacuum probability.  We consider both these measures, and find quench
protocols $Q(t)$ that minimize the excess energy and those that maximize the vacuum-to-vacuum
probability.

We first show that both the excess energy and the overlap depend only on the occupancies of bosonic
modes at the end of the quench. 
We find that the derivative of the optimal protocol must be symmetric with respect to
$\tau/2$, and the protocol function itself must obey $Q(t)=1-Q(\tau-t)$. 

The shape of the finite-duration quench is parametrized as a Fourier series, and its coefficients
are optimized.  Fast protocols excite high energy modes, and thus are non-universal in the Luttinger
liquid sense.  With increasing $\tau$, the excess energy is minimized by the a smooth protocol while
the overlap is maximized by a linear ramp.  
In this limit, the minimal energy and maximal overlap are both universal functions of the system size and the duration of the protocol.


In Section \ref{sec:model}, we first introduce the model, the quench protocol, and notations, and
then derive expressions for the excess energy and for the overlap with the final ground state.  The
parity of the optimal quench protocol is considered in Section \ref{sec:parity}.  In Sections
\ref{sec:minimize_energy} and \ref{sec:maximize_overlap} we report on the the optimization of $Q(t)$
by respectively minimizing the final energy and maximizing the final overlap with the target state.
Section \ref{sec:conclusion} provides some concluding discussion.

\section{Quantum quench in the Luttinger model} \label{sec:model}

The low-energy behaviour of one-dimensional electron system is described by the Luttinger model.
This model has the advantage that both the non-interacting and the interacting system can be
diagonalized analytically. This is because both the kinetic and the interaction energy can be
expressed as quadratic terms of bosonic creation and annihilation operators describing electron-hole
excitations. In this paper, a quantum quench from the non-interacting to the interacting Luttinger
model is considered in such a way that the system is prepared into the ground state of the
non-interacting system initially. The time dependent Hamiltonian is given as
\begin{gather}
H(t)=H_{0}+Q(t)V
\label{eq:tdham}
\end{gather}
where 
\begin{gather}
H_{0}=\sum_{q>0}\omega_{0}(q)\left(b^{+}_{q}b_{q}+b^{+}_{-q}b_{-q}\right)
\end{gather} 
is the Hamiltonian of the non-interacting system with $\omega_{0}(q)=v|q|$. In the formula $b_q$ is the bosonic annihilation operator corresponding to the wavenumber $q$. The second term in Eq. \eqref{eq:tdham} describes the electron-electron interaction
\begin{gather}
V=\sum_{q>0}g(q)\left(b^{+}_{q}b^{+}_{-q}+b_{q}b_{-q}\right)
\end{gather}
where $g(q)=g_{2}|q|e^{-v\tau_{0}|q|}$. Note that in the interaction, only back-scattering ($g_2$) is considered. It can be shown that the forward scattering ($g_4$) does not effect the bosonic occupation numbers to leading order in the interaction strength and, hence, can be neglected. The time scale of $\tau_0$ is introduced to model the high energy cut-off and is assumed to be inverse proportional to the bandwidth of the electron system.

In Eq. \eqref{eq:tdham}, $Q(t)$ describes the quench protocol with the duration of $\tau$, i.e., 
\begin{gather}
Q(t)=\left\{\begin{array}{cc} 0 & \mbox{if $t<0$} \\ Q(t) & \mbox{if $0<t<\tau$} \\
1 & \mbox{if $t>\tau$}\end{array}\right.
\end{gather}
where the non-trivial time dependence happens in the intermediate interval.

If the quench is adiabatic, i.e., in the $\tau\rightarrow\infty$ limit, the system is expected to arrive in the ground state of the interacting system after the quench and no bosonic excitations are present. However, if the quench duration is finite, the final state is presumably not the pure ground state of the interacting Hamiltonian but is a linear combination of the ground state and excited states.

The bosonic excitations of the interacting system are described by the operators of
\begin{gather}d_{\pm q}=b_{\pm q}\sqrt{\frac{\omega_{0}(q)}{2\Omega(q)}+\frac{1}{2}}\,\,+\,\,b_{\mp q}^{+}\sqrt{\frac{\omega_{0}(q)}{2\Omega(q)}-\frac{1}{2}}
\end{gather}
which diagonalize the interacting Hamiltonian as 
\begin{gather}
H(\tau)=E_{\mathrm{GS}}+\sum_{q>0}\Omega(q)\left(d^{+}_{q}d_{q}+d^{+}_{-q}d_{-q}\right)
\label{eq:htau}
\end{gather}
where $E_{\mathrm{GS}}=\sum_{q>0}\left(\Omega(q)-\omega_{0}(q)\right)$
is the ground state energy and $\Omega(q)=\sqrt{\omega_{0}(q)^{2}-g(q)^{2}}$
is the spectrum of the elementary excitations.

The dynamics during the quantum quench may be described by the time dependent annihilation operators as
\begin{gather}
b_{q}(t)=u_{q}(t)b_{q}+v_{q}^{*}(t)b_{-q}^{+}\nonumber \\b_{-q}(t)=u_{q}(t)b_{-q}+v_{q}^{*}(t)b_{q}^{+} 
\end{gather}
where the coefficients obey
\begin{gather}
\label{eq:tdschr}
i\hbar\partial_{t}\left[\begin{array}{c} u_{q}(t) \\ v_{q}(t) \end{array}\right]=\left[\begin{array}{cc} \omega_{0}(q) & Q(t)g(q) \\ -Q(t)g(q) & -\omega_{0}(q) \end{array}\right]\left[\begin{array}{c} u_{q}(t) \\ v_{q}(t) \end{array}\right]
\end{gather}
with the initial conditions $u_{q}(0)=1$ and $v_{q}(0)=0$. At any time instant $|u_{q}(t)|^{2}-|v_{q}(t)|^{2}=1$ holds true.

By means of the $u_q(t)$ and $v_q(t)$ coefficients, the time-dependent wavefunction is expressed as
\begin{gather}
|\Psi(t)\rangle=\prod_{q>0}\left[\frac{1}{u_q^*(t)}\mathrm{exp}\left(i\omega_0(q)t+\frac{v^{*}_q(t)}{u^{*}_q(t)}b_q^{+}b_{-q}^{+}\right)\right]|0\rangle
\label{eq:psi}
\end{gather}
where $|0\rangle$ is the initial ground state of the non-interacting system \cite{balazs_loschmidt}. The wavefunction depends on the protocol function $Q(t)$ through the coefficients $u_q(t)$ and $v_q(t)$.

In the present paper, our main goal is to study the optimal $Q(t)$ protocol function with finite duration $\tau$ which results in a final state $|\Psi(\tau)\rangle$ closest to the ground state of the 
interacting Hamiltonian $H(\tau)$. We investigate two different quantities which both represent a measure of how far the final state is from the interacting ground state. 
One of them is the expectation value of the total energy in the final state $E_f = \langle\Psi(\tau)|H(\tau)|\Psi(\tau)\rangle$. The other quantity is the overlap between the time evolved final state and the 
ground state of the interacting system $P_{\mathrm{GS}}=|\langle \mathrm{GS}|\Psi(\tau)\rangle|^{2}$. In other words, $P_\mathrm{GS}$ is the transition probability from the non-interacting to the interacting vacuum. 
Note that this quantity has been considered numerically in Ref. \onlinecite{rahmani_prl2011} as the measure for optimization in a related problem.

Our aim is to find the optimal protocol function which minimizes $E_f$ or
maximizes $P_\mathrm{GS}$. These two quantities are represented
as functionals of $Q(t)$.

For generic quench protocol, the energy functional is obtained by calculating the expectation value of Eq. \eqref{eq:htau} as
\begin{gather}
E_f[Q] = E_\mathrm{GS} +\sum_{q>0}2\Omega(q)n_q[Q] 
\label{eq:ef}
\end{gather}
where the occupation number
is the expectation value of the boson numbers in the $+q$ or $-q$ channel. The occupation number
\begin{gather}
n_{q}[Q]=\langle d_{\pm q}^{+}d_{\pm q}^{}\rangle = \frac{\omega_{0}(q)}{2\Omega(q)}\left(|u_{q}(\tau)|^{2}+|v_{q}(\tau)|^{2}\right)+\nonumber\\
+\frac{g(q)}{2\Omega(q)}\left(u_{q}(\tau)^{*}v_{q}(\tau)+u_{q}(\tau)v_{q}(\tau)^{*}\right)-\frac{1}{2}
\label{eq:nq}
\end{gather}
depends on the protocol function through the coefficients $u_q(t)$ and $v_q(t)$.

The vacuum-to-vacuum probability is obtained by taking the overlap of Eq. \eqref{eq:psi} with the ground state of the interacting system. Interestingly, the probability depends on the protocol function again through the occupation number only as
\begin{gather}
\ln P_{\mathrm{GS}}[Q]=-\sum_{q>0}\ln\left(1+n_{q}[Q]\right)\,.
\label{eq:pgs}
\end{gather}

The functionals $E_f[Q]$ and $P_\mathrm{GS}[Q]$ are highly non-linear in the protocol function and finding the optimum for arbitrary interaction strength is very complicated using analytic methods. Therefore, the following discussion is restricted to the limiting case of weak interactions. To leading order in the perturbation theory, i.e., when $g_2\ll v$ holds, the occupation number is given by
\begin{gather}
n_{q}[Q]=\frac{g(q)^{2}}{4\omega_{0}(q)^{2}}\left|\int_{0}^{\tau}\mathrm{d}t\,Q'(t)e^{2i\omega_{0}(q)t}\right|^{2}
\label{eq:nqofQ}
\end{gather}
where $Q'(t)$ is the derivative of the quench protocol. It can be shown that even if forward scattering ($g_4$) were considered in the interacting Hamiltonian the leading term in Eq. \eqref{eq:nqofQ} would not depend on $g_4$.
We substitute Eq. \eqref{eq:nqofQ} into Eqs. \eqref{eq:ef} and \eqref{eq:pgs} and keep terms to leading order in the perturbation.
In the thermodynamic limit, the summation over the wavenumbers turns into an integral leading to
\begin{gather}
\varepsilon_f[Q]= \frac{E_{f}[Q]-E_\mathrm{GS}}{|E_\mathrm{GS}|}=\nonumber \\
=\int_{0}^{\tau}\mathrm{d}t\int_{0}^{\tau}\mathrm{d}t'Q'(t)Q'(t')\frac{\tau_{0}^{2}(\tau_{0}^{2}-(t-t')^{2})}{(\tau_{0}^{2}+(t-t')^{2})^{2}}
\label{eq:efunc}
\end{gather}
with the ground state energy of
\begin{gather}
E_{\mathrm{GS}}=-\frac{L}{16\pi v\tau_{0}}\left(\frac{g_{2}}{v}\right)^{2}\frac{1}{\tau_{0}}
\label{eq:gs}
\end{gather}
and
\begin{gather}
\mathcal{F}[Q] = \frac{\ln P_{\mathrm{GS}}[Q]}{|E_{\mathrm{GS}}|\tau_{0}} = \nonumber \\
=-\int_{0}^{\tau}\mathrm{d}t\int_{0}^{\tau}\mathrm{d}t'Q'(t)Q'(t')\frac{\tau_{0}^{2}}{\tau_{0}^{2}+(t-t')^{2}}
\label{eq:pfunc}
\end{gather}
where the dimensionless and non-extensive quantities of $\varepsilon_f$ and $\mathcal{F}$ have been introduced. In Eq. \eqref{eq:gs}, $L$ is the length of the system which is considered to be in the thermodynamic limit. 

In the following sections, our goal is to find the protocol function $Q(t)$ which minimizes $\varepsilon_f$ or maximizes $\mathcal{F}$.
We note that the formulae in Eqs. \eqref{eq:efunc} and \eqref{eq:pfunc} are valid to leading order of the perturbation theory which is maintained as long as $n_{q}[Q]$ is small in all momentum modes.

\section{Parity of the optimal quench protocol}
\label{sec:parity}

An important feature of the optimal quench protocol is its symmetries, e.g., the parity.   If the
protocol is known to have a symmetry, this could reduce significantly the (numerical) effort in
determining the optimal ramp. 

In Eqs. \eqref{eq:efunc} and \eqref{eq:pfunc}, we observe that functionals depend on the derivative
of the protocol function. Let us split up the derivative of the protocol function into even and odd
part as
\begin{gather}
Q'(t)= p(t) = p_a(t) + p_s(t)
\end{gather}
where $p_a(t) = (Q'(t)-Q'(\tau-t))/2$ is the anti-symmetric part while $p_s(t) = (Q'(t)+Q'(\tau-t))/2$ is the symmetric part. The boundary conditions of the protocol function demand
\begin{gather}
\int_0^{\tau} p_s(t)\,\mathrm{d}t = 1
\label{eq:ptcond}
\end{gather}
but the anti-symmetric part can be an arbitrary, odd function since its integral vanishes on $[0,\tau]$.

In both the energy functional and the vacuum-to-vacuum transition probability, the kernel of the integral is symmetric under $t\rightarrow \tau-t$ and $t'\rightarrow \tau -t'$, i.e., when both time variables are reflected. Therefore, the integral part of the functionals are rewritten as
\begin{gather}
\int_{0}^{\tau}\mathrm{d}t\int_{0}^{\tau}\mathrm{d}t'p_s(t)p_s(t')K(t-t') +\nonumber \\ + \int_{0}^{\tau}\mathrm{d}t\int_{0}^{\tau}\mathrm{d}t'p_a(t)p_a(t')K(t-t')
\label{eq:functional}
\end{gather}
and the cross terms proportional to the integral of $p_s(t)p_a(t')$ vanish. $K(t)$ is the kernel of Eqs. \eqref{eq:efunc} or \eqref{eq:pfunc}, respectively.

The kernel of the integral is positive (negative) definite for the final energy $\varepsilon_f$ (transition probability $\mathcal{F}$). This is because the total energy is bounded from below by the ground state energy and the probability $P_\mathrm{GS}$ is bounded from above by 1.
In principle, the boundedness would allow semi-definite kernels but it can be proven by means of Fourier transformation that the kernels of \eqref{eq:efunc} and \eqref{eq:pfunc} have no zero eigenvalue on the space of functions with finite duration.

As a consequence, the kernels are positive (negative) definite and so are they on the subspaces of even and odd functions separately.
Therefore, the second term of (\ref{eq:functional}) is minimized (maximized) by $p_a(t)=0$. In the first term, the symmetric part cannot be chosen as an identically zero function because it would not satisfy the boundary condition Eq. \eqref{eq:ptcond}. 
For the anti-symmetric part, however, no such condition is prescribed.

Thus, $p_a(t)=0$ minimizes the second integral in Eq. \eqref{eq:functional}, which means that the optimal $Q'(t)$ function must be an even function, i.e., symmetric under the reflection of $t\rightarrow \tau -t$. Consequently, $Q(t)=1-Q(\tau-t)$ for the optimal quench.
In the following sections, protocol functions with this symmetry property will be considered only.

\section{Optimal quench minimizing the final energy}  \label{sec:minimize_energy}

\begin{figure*}[t!]
\centering
a) \quad\includegraphics[width=7.8cm]{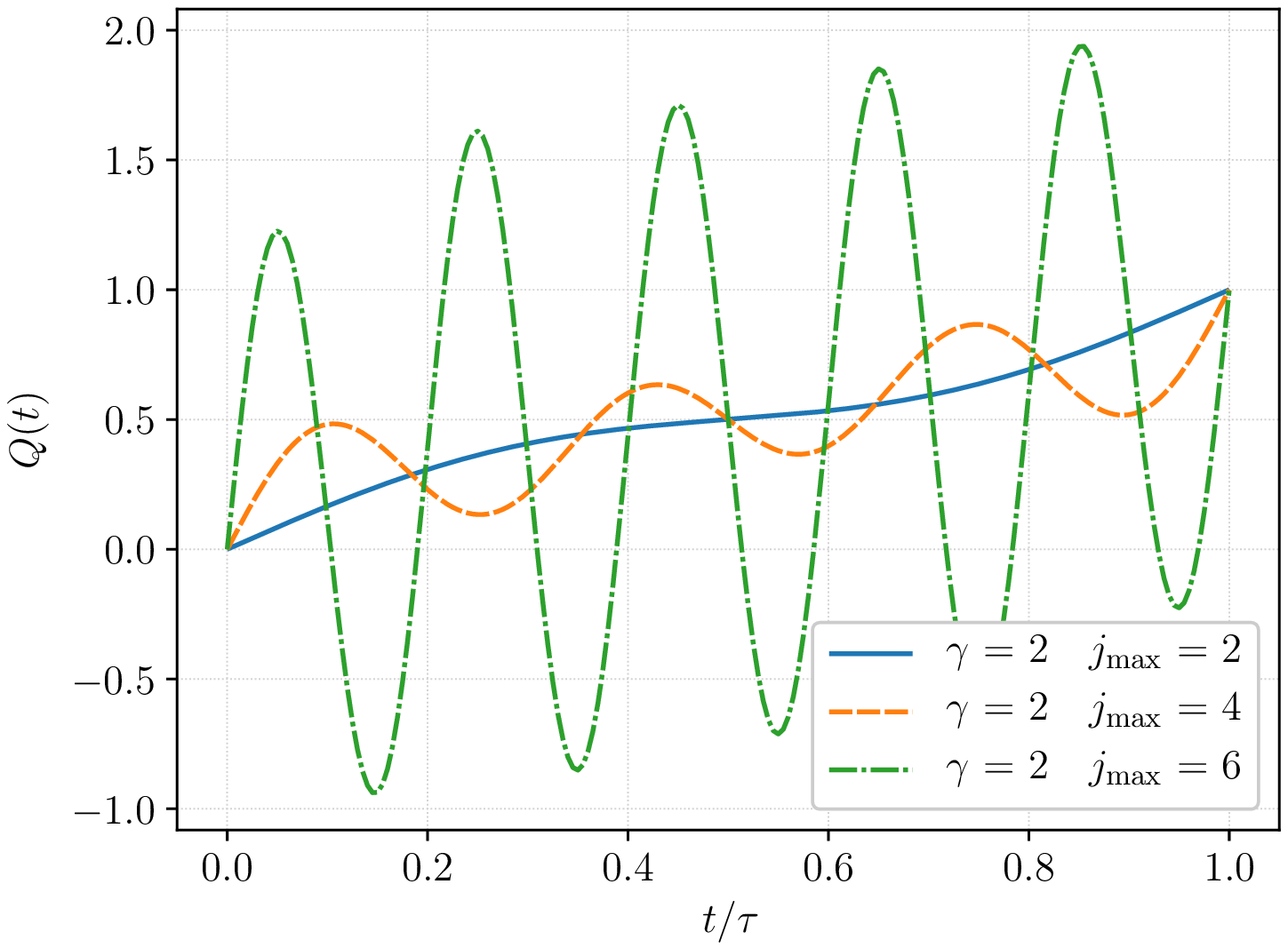}\qquad b) \quad
\includegraphics[width=7.8cm]{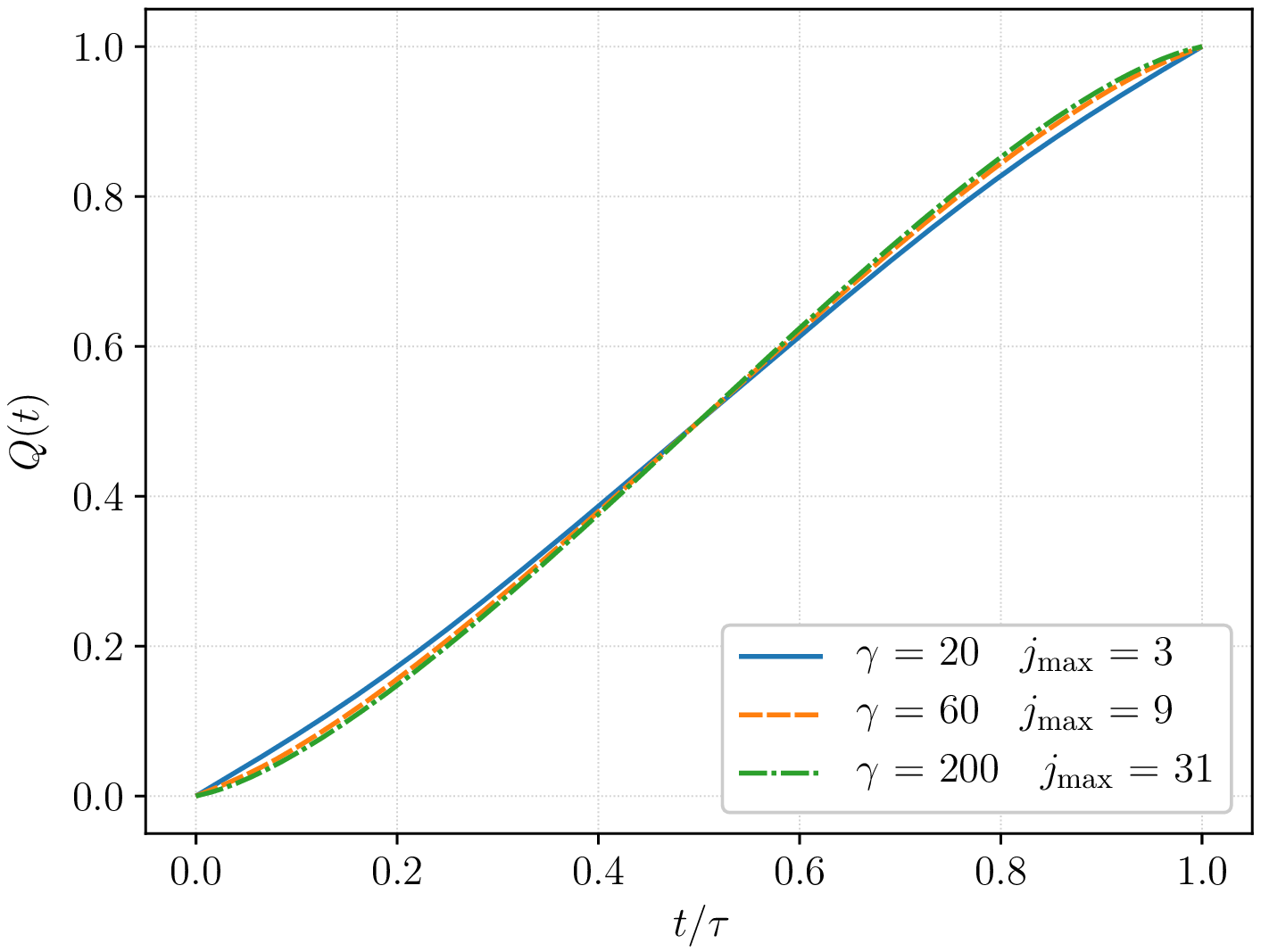}
\caption{a) The optimal quench for $\gamma = 2$ with different values of $j_\mathrm{max}$, the
  number of Fourier modes retained.  Increasing the truncation index $j_\mathrm{max}$ leads to more
  oscillating behavior of the optimal quench. b) The optimal quench minimizing the final energy for
  different values of $\gamma$ and with $j_\mathrm{max}\approx \gamma/(2\pi)$.} 
\label{fig:gamma1}
\end{figure*}

This section focuses on minimizing the final energy $\varepsilon_f[Q]$ as defined in Eq. \eqref{eq:efunc}.

Let us consider the Fourier expansion of $Q'(t)$ as
\begin{gather}
Q'(t)=\sum_{j=0}^{\infty}\frac{a_j}{\tau}\cos(\omega_j t)
\label{eq:qpt_fourier}
\end{gather}
where the frequencies $\omega_j=2\pi j/\tau$ have been introduced. Note that the Fourier expansion does not involve any sine function since even functions are considered only in accordance with Sec. \ref{sec:parity}. By using the Fourier expansion, our goal is to find the optimal coefficients $a_j$.
The final energy functional is obtained as
\begin{gather}
\varepsilon_f[Q] = \sum_{j,j'=0}^{\infty}a_j M_{jj'} a_j'
\label{eq:en1}
\end{gather}
where the matrix elements of $\mathbf{M}$ are defined as
\begin{gather}
M_{jj'}=\frac{1}{\gamma^2}\int_0^1\mathrm{d}x\,\int_0^1\mathrm{d}x'\,\cos\left(2\pi j x\right)\cos\left(2\pi j' x'\right)\times\nonumber\\
\times \dfrac{\dfrac{1}{\gamma^2}-(x-x')^{2}}{\left(\dfrac{1}{\gamma^2}+(x-x')^{2}\right)^{2}}
\end{gather}
with
\begin{gather}
\gamma = {\tau}/{\tau_0}
\end{gather}
being the dimensionless quench duration.
For the $a_0$ coefficient, $a_0=1$ must hold which ensures that the integral of $Q'(t)$ is one.
This condition and the minimization of Eq. \eqref{eq:en1} result in the optimal coefficients of
\begin{gather}
a_{\mathrm{opt},j} = \varepsilon_\mathrm{min}\cdot \left(\mathbf{M}^{-1}\right)_{j1}
\label{eq:aopt}
\end{gather}
where $\varepsilon_\mathrm{min} = 1/\left(\mathbf{M}^{-1}\right)_{11}$ is the minimal energy.

The matrix elements of $M_{jj'}$ cannot be expressed in a closed form for any $j$ and $j'$. Therefore, numerical integration is applied. For the numerical calculation, the Fourier series is truncated at $j_\mathrm{max}$, i.e., only Fourier components from $j=0$ to $j=j_\mathrm{max}-1$ are allowed. Then, the matrix $\mathbf{M}$ has the size of $j_\mathrm{max}\times j_\mathrm{max}$. In the simulation, the optimal coefficients are computed based on Eq. \eqref{eq:aopt} and the optimal protocol function is reconstructed based on Eq. \eqref{eq:qpt_fourier}.

Let us first study shorter quenches, for example $\gamma=2$. The numerically computed optimal quench is shown in Fig. \ref{fig:gamma1} a) for different values of $j_\mathrm{max}$. As the truncation index $j_\mathrm{max}$ increases, the optimal quench exhibits oscillations with larger and larger amplitude. If further Fourier components are allowed in the quench protocol, the optimal protocol function becomes even more oscillating with even larger amplitudes. These high frequency components with large amplitude excite bosons far beyond the cutoff energy $1/\tau_0$. In this regime, however, the linear spectrum of the Luttinger model does not apply anymore and, hence, the highly oscillating optimal quench is the consequence of unphysical effects.

In order to stay inside the validity of the Luttinger model, we allow Fourier components with frequencies up to the cutoff energy, i.e. $\omega_j\lesssim 1/\tau_0$. In terms of $j$ indices, $j\lesssim \gamma/(2\pi)$ must hold which means that $j_\mathrm{max}$ should be chosen around $\gamma/(2\pi)$. This also implies that quenches shorter than $2\pi\tau_0$ inevitably generate excitations in the high energy regime and, hence, are beyond the validity of the Luttinger model independently from the quench protocol function.

Fig. \ref{fig:gamma1} b) shows optimal quench protocol functions in which the truncation index $j_\mathrm{max}$ is chosen as the integer part of $\gamma/(2\pi)$. With this truncation, the optimal protocols are found to be non-oscillating, smooth functions.

Numerical results indicate that the optimal protocol function converges when the quench duration reaches the range of $100\tau_0$. In this regime, it is also observed in the simulation that increasing $j_\mathrm{max}$ does not effect the optimal quench protocol and neither leads to oscillations.
It is an interesting question, how the limiting protocol function can be expressed analytically.

The long quench limit of the functional in Eq. \eqref{eq:efunc} is calculated as
\begin{gather}
\varepsilon_f[Q] = \left[\left(\tau Q'(0)\right)^2+\left(\tau Q'(\tau)\right)^2\right] \frac{\ln\gamma}{\gamma^2} + \mathcal{O}\left(\gamma^{-2}\right)
\label{eq:leading}
\end{gather}
if the protocol function is an analytic function of time. Note that $Q'$ scales with $\tau^{-1}$, therefore, the leading term is proportional to $\tau^{-2}\ln(\tau/\tau_0)$. 
Interestingly, the leading term of the energy functional depends on the derivative of the protocol function evaluated only at the edges of the quench interval. To minimize the leading term in Eq. \eqref{eq:leading}, the optimal protocol function must fulfill 
\begin{gather}
Q'(0)=Q'(\tau)=0
\label{eq:longq_cond}
\end{gather}
for long quenches. The exact characteristics of $Q(t)$ is then chosen in such a way that next-to-leading corrections are minimized.

Since this problem is complicated using analytical methods, numerical method is applied.
During the simulation it is found that for a long quench duration, the optimal Fourier coefficients obey power-law behavior as
\begin{gather}
a_{\mathrm{opt},j} = \left\{ \begin{array}{cc} 1 & \mbox{if $j=0$} \\ -\frac{A}{j^\beta} & \mbox{if $j\geq 1$} \end{array}\right.
\label{eq:powercoeffs}
\end{gather}
where $A$ and $\beta$ are numeric parameters.
The power-law behavior is also expected for long quenches when the energy scales $1/\tau$ and $1/\tau_0$ are widely separated. Between these scales, there is a wide energy range in which no dominant energy scale is present and, hence, the $a_j$ coefficients are expected to obey a scale-free $j$-dependence.

\begin{figure*}[t!]
\centering
a)\quad\includegraphics[width=8cm]{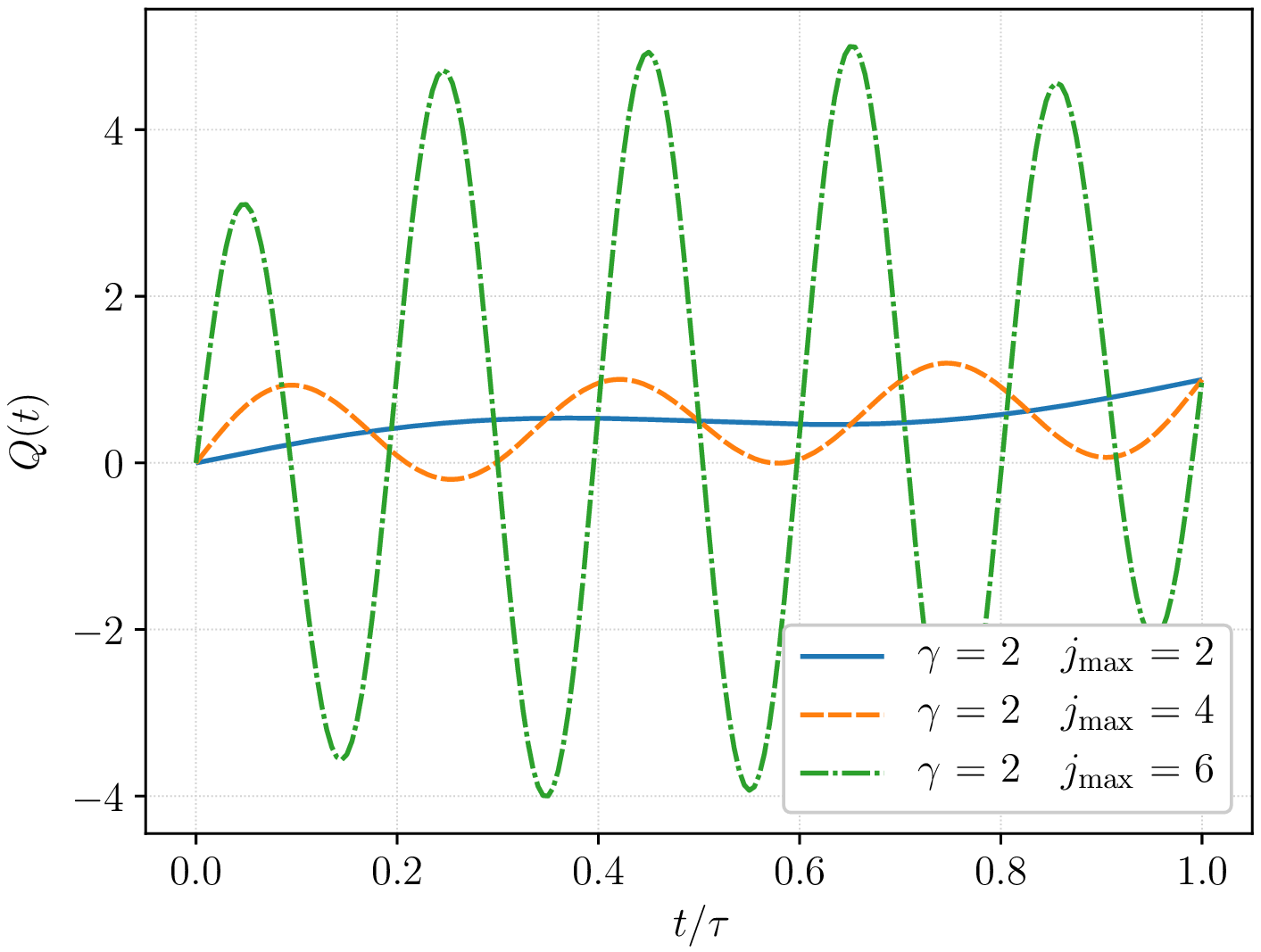}\qquad b)\quad
\includegraphics[width=8cm]{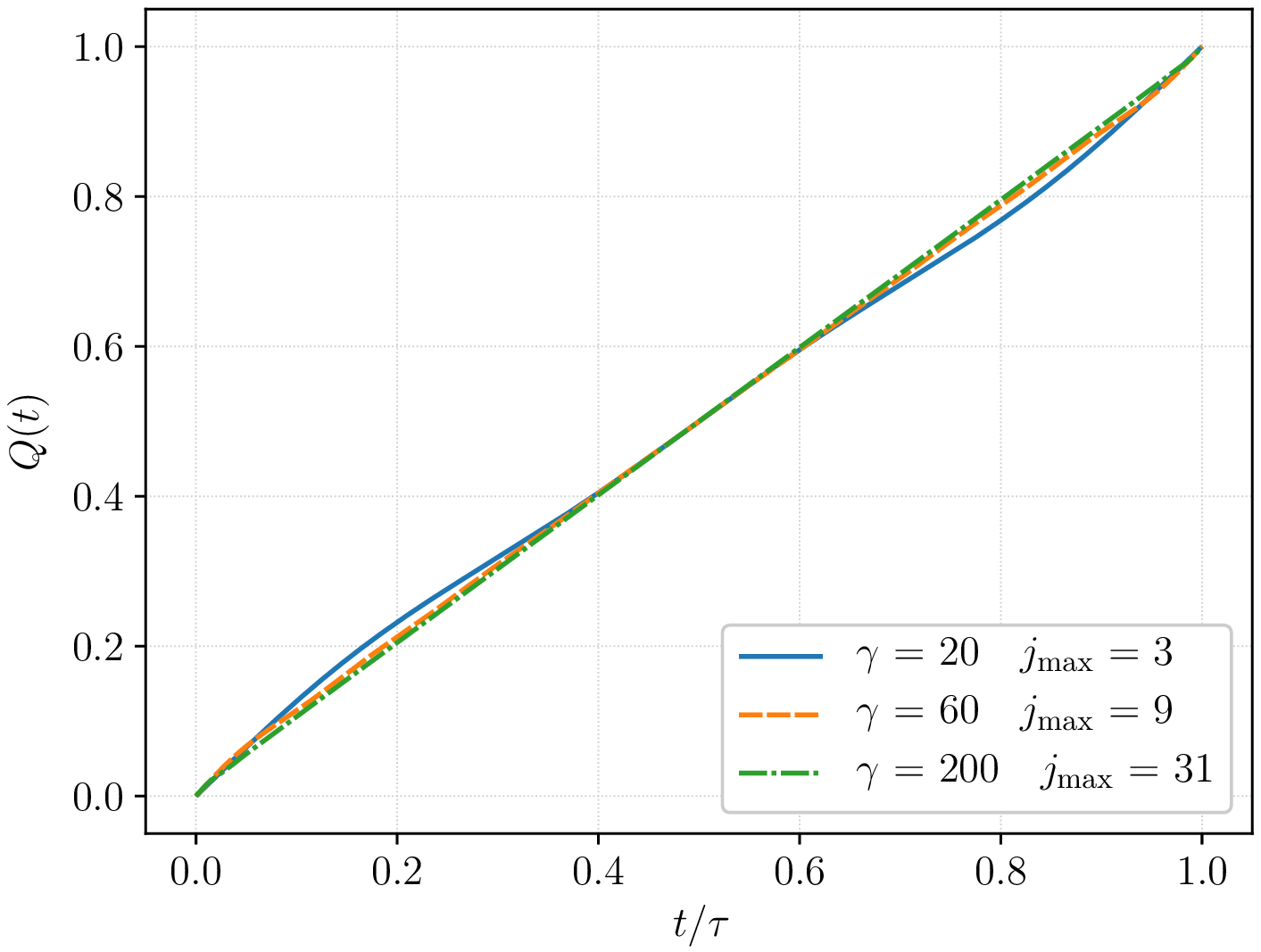}
\caption{a) The optimal quench for $\gamma = 2$ with different values of the truncation index
  $j_\mathrm{max}$.  For larger $j_\mathrm{max}$, the optimal quench has more oscillations.  b) The
  optimal quench maximizing the vacuum-to-vacuum transition probability for different values of
  $\gamma$ and with $j_\mathrm{max}\approx \gamma/(2\pi)$. }
\label{fig:gammas_lnpgs}
\end{figure*}

Applying the condition of \eqref{eq:longq_cond} to the protocol function with power-law Fourier components described in Eq. \eqref{eq:powercoeffs}, 
\begin{gather}
Q'(0)=\frac{1}{\tau}\left(1-\sum_{j=1}^{\infty}\frac{A}{j^\beta}\right) = 0 \Longrightarrow A = \frac{1}{\zeta(\beta)}
\end{gather}
where $\zeta$ is the Riemann zeta function.
Therefore, the optimal protocol function reads as
\begin{gather}
Q(t) = \frac{t}{\tau}- \frac{1}{2\pi\zeta(\beta)}\sum_{j=1}^{\infty}\frac{\sin(\omega_j t)}{j^{\beta+1}} = \nonumber \\
= \frac{t}{\tau} - \frac{1}{2\pi\zeta(\beta)}\mathrm{Im}\left\{\mathrm{Li}_{\beta+1}\left(e^{i2\pi\frac{t}{\tau}}\right)\right\}
\label{eq:optquench_energy}
\end{gather}
with Li$_\nu(x)$ being the polylogarithm function.
The value of the $\beta$ parameter must be set in such a way that the next-to-leading term in the energy functional proportional to $\gamma^{-2}$ is minimized.
The optimal value cannot be obtained analytically but must be handled numerically. We performed simulations with durations up to $\tau/\tau_0=10000$ and where the truncation index ranges from 10 to 50. The $\beta$ parameter is obtained by non-linear curve fit on the optimal $a_j$ coefficients. 

Based on the numerical study, the optimal quench of a long duration has the form of Eq. \eqref{eq:optquench_energy} with approximately 
\begin{gather}
\beta\approx 1.4\,.
\end{gather}
The minimal energy is approximately $\varepsilon_\mathrm{min}\approx 8.0 \cdot\gamma^{-2}$ and
\begin{gather}
E_\mathrm{min} = E_\mathrm{GS} + 8.0 \left(\frac{g_2}{v}\right)^2 \frac{L}{16\pi v\tau^2}\,.
\label{eq:emin}
\end{gather}
The second term measures the energy amount which is inevitably present in the form of excitations after a finitely long quench. Interestingly, this term is independent from the cutoff $1/\tau_0$ and is, therefore, universal for one-dimensional systems within perturbation theory.

Finally, we note that for short times, the optimal quench protocol behaves as a power-law function with the exponent of $\beta$ as
\begin{gather}
Q(t\ll \tau)\approx \frac{\sin\left(\frac{\beta\pi}{2}\right)\Gamma(-\beta)}{2\pi\zeta(\beta)}\left(2\pi\frac{t}{\tau}\right)^{\beta}
\end{gather}
where $\Gamma(x)$ is the gamma function.

\section{Optimal quench maximizing the vacuum-to-vacuum transition probability}  \label{sec:maximize_overlap}

In this section, the optimal quench maximizing the overlap between the final state and the interacting ground state as defined in Eq. 
\eqref{eq:pfunc} is studied.
Similarly to the final energy, the Fourier series of $Q'(t)$ is considered as given in Eq. \eqref{eq:qpt_fourier}. Numerical results imply that Fourier components with frequencies above the cutoff $1/\tau_0$ result in unphysical oscillations for shorter quenches. In order to stay within the validity of the Luttinger model, Fourier components above the cutoff should be omitted and, therefore, the truncation index $j_\mathrm{max}$ is chosen as the integer part of $\gamma/(2\pi)$ for the numerical simulation.

Numerical results are shown in Fig. \ref{fig:gammas_lnpgs}. These indicate that the optimal quench tends to be linear for longer quenches. In the case of $\tau\gg\tau_0$, the optimal quench can be derived analytically. First, the functional $\mathcal{F}$ is rewritten as
\begin{gather}
\mathcal{F}[Q] = -\frac{\pi}{\gamma} \int_{0}^{\tau}\mathrm{d}t\,\int_{0}^{\tau}\mathrm{d}t'\,Q'(t)Q'(t')\delta_{\gamma}\left(\frac{t-t'}{\tau}\right)
\end{gather}
where
\begin{gather}
\delta_\gamma(x)=\frac{1}{\pi}\frac{\frac{1}{\gamma}}{\frac{1}{\gamma^2}+x^2}
\end{gather}
has been introduced.
In the limit of long quench, $\lim_{\gamma\rightarrow\infty}\delta_{\gamma}(x)=\delta(x)$ is the Dirac-delta function and the functional is obtained as
\begin{gather}
\mathcal{F}[Q] = -\frac{\pi}{\gamma}\int_{0}^{\tau}\mathrm{d}t\,\tau\left(Q'(t)\right)^{2}\,.
\end{gather}
This functional is maximized by the linear quench
\begin{gather}
Q(t)=\frac{t}{\tau}\,.
\label{eq:optquench_lnpgs}
\end{gather}
Interestingly, this optimal, linear quench is the $\beta\rightarrow 1$ limit of the optimal quench for the minimal energy given in Eq. \eqref{eq:optquench_energy}. The maximal probability is calculated as $\mathcal{F}_\mathrm{max}= -\pi/\gamma$ and, hence,
\begin{gather}
\ln P_{\mathrm{GS,max}}=-\left(\frac{g_{2}}{v}\right)^{2}\frac{L}{16v\tau}
\label{eq:fmax}
\end{gather}
which is also a universal value since independent from the cut-off, $\tau_{0}^{-1}$. Note that Eq. 
\eqref{eq:fmax} describes the maximal probability of finding the final state in the interacting ground state if a quantum quench of finite duration is applied.

\section{Summary \& Discussion}   \label{sec:conclusion}


\begin{figure}[h]
\centering
\includegraphics[width=8cm]{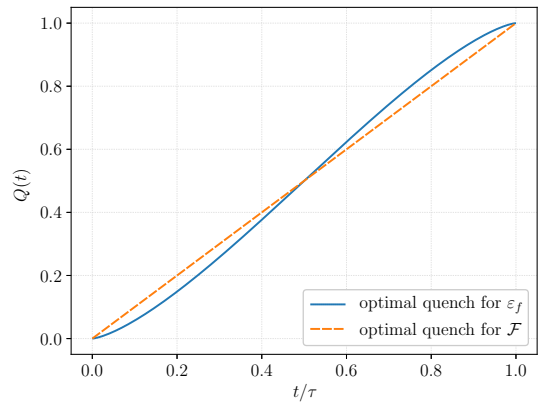}
\caption{The optimal protocol function which minimizes the final energy (blue/dark) from
  Eq. \eqref{eq:optquench_energy}, and which maximizes the vacuum-to-vacuum probability
  (orange/light).  The latter  has a simple linear time dependence.}
\label{fig:summary}
\end{figure}

In this work, we studied the non-equilibrium behavior of the Luttinger model under finite-rate
quenches.  The low energy bosonic Hamiltonian in Eq. \eqref{eq:tdham} depends on time through the
protocol function $Q(t)$ which switches on a weak interaction.  We optimized the $Q(t)$ so as to get
the system as close to the ground state of the final Hamiltonian as possible by the end of the
quench.  Two measures of deviation from the target state were used for this purpose: the excess
energy at the end of the quench and the overlap between the time evolved final wavefunction and the
interacting ground state.

We have shown that the optimal protocol must be symmetric with respect to the midpoint of the quench
duration.  For short quenches $\tau\lesssim\tau_0$, the optimal quench exhibits sharp oscillations
which are related to bosons excited to very high energies and are beyond the realm of the effective
low energy model.  To avoid these excitations, longer quenches with $\tau\gg\tau_0$ are considered within the validity range of 
Luttinger model. In this case, the optimal quenches do not exhibit wild oscillation and the protocol functions are found in closed forms in Eqs. 
\eqref{eq:optquench_energy} and \eqref{eq:optquench_lnpgs} for the case of weak final
interactions. The optimal protocols are shown in Fig. \ref{fig:summary}.  

For these ramp protocols, the minimal energy and the maximal vacuum-to-vacuum probability are
expressed in Eqs. \eqref{eq:emin} and \eqref{eq:fmax}. These values are independent of the cut-off
and are therefore universal within the perturbation theory.  These analytical protocol functions,
shown in Fig. \ref{fig:summary}, are optimal in the thermodynamic limit. It remains to be
investigated to what extent these protocols remain valid beyond the realm of weak interactions.

Our approach of expanding in a Fourier series up to a physically motivated cutoff is different in
spirit from finding numerically exact optimum paths by large-scale numerics (e.g., as done in
Refs.\ \onlinecite{Caneva_Calarco_Fazio_optimal_PRL2009,rahmani_prl2011}), or from finding
mathematically exact optimal protocols for systems having a simpler description (e.g., as done in
Ref.\ \onlinecite{DelCampo_Muga_PRL2010, Takahashi_PRE2013}).  One could also expand $Q(t)$ in a
power series; we have found that the same main results (oscillatory $Q(t)$ for small $\tau$;
different universal curves for minimizing energy and for maximizing overlap) are also found with
such an expansion.  However, we believe that the Fourier description presented in this paper has a
more physical interpretation.

\begin{acknowledgments}
This research is supported by the National Research, Development and Innovation Office - NKFIH   within the Quantum Technology National Excellence Program (Project No.
      2017-1.2.1-NKP-2017-00001), K119442, SNN118028, by the  BME-Nanonotechnology FIKP grant of EMMI (BME FIKP-NAT)  and by
 UEFISCDI, project number PN-II-RU-TE-2014-4-0432.
\end{acknowledgments}

\bibliographystyle{apsrev}
\bibliography{optramp}

\end{document}